# Urban volumetrics: spatial complexity and wayfinding, extending space syntax to three-dimensional space

Lingzhu ZHANG[1]

[1] College of Architecture and Urban Planning, Tongji University,
zhanglz@tongji.edu.cn,
College of Architecture and Urban Planning, Tongji University, 1239 Siping Road, Yangpu District, Shanghai, China

Alain J F CHIARADIA[2]*
*Corresponding Author
[2] Faculty of Architecture, Department of Urban Planning and Design, The University of Hong Kong, alainjfc@hku.hk, Tel: +852 3917 6169
Room 814, 8/F, Knowles Building, Department of Urban Planning and Design, the University of Hong Kong, Pokfulam Road, Hong Kong  SAR

## Abstract

Wayfinding behavior and pedestrian movement pattern research relies on objective spatial configuration representation and analysis, such as space syntax, to quantify and control for the difficulty of wayfinding in multi-level buildings and urban built environments. However, the space syntax's representation oversimplifies multi-level vertical connections. The more recent segment and angular approaches to space syntax remain un-operationalizable in three-dimensional space. The two-dimensional axial-map and segment map line representations are reviewed to determine their extension to three-dimensional space line representation. Using an extreme case study research strategy, four representations of a large scale complex multi-level outdoor and indoor built environment are tested against observed pedestrian movement patterns N = 17,307. Association with the movement patterns increases steadily as the representation increases toward high three-dimensional space level of definition and completeness. A novel hybrid angular-Euclidean analysis was used for the objective description of three-dimensional built environment complexity. The results suggest that pedestrian wayfinding and movement pattern research in a multi-level built environment should include interdependent outdoor and indoor, and use full three-dimensional line representation.

# Introduction

During walking navigation and wayfinding, cognitive representation of the Built Environment (BE), stored information, relative knowledge relation, and spatial feature attributes are continuously updated (Montello, 2005). Various spatial representations and metrics (Boeing, 2019; Coutrot, et al., 2020) are used, such as entropy, to objectively characterize the BE in relation to spatial complexity and wayfinding, linking objective measures of space to cognition and action (Zimring & Dalton, 2003). This study focuses on objective spatial representations of BE layout complexity used in multi-level BE pedestrian movement pattern and wayfinding studies.

Objective characterization of the BE can be broadly categorized into two classes: spatial composition and spatial configuration. Spatial composition is quantifying the presence, absence, and nature of spatial features such as signs, parks, restaurants, lobbies, and street direction. Spatial configuration focuses on the relationship between spatial features (Small & Adler, 2019). A spatial configurational description enables the analysis and comparison of the individual elements of the composition, generic nature of the configuration for outdoor and indoor BE (Hillier & Hanson, 1984; Hillier, 1996; Hanson, 1998), their interdependencies from different spatial positions, and how the configuration is changed when the composition or relation is changed, i.e., how the part can affect the whole. For a given composition description size, there is a vast number of possible configuration combinations (Hillier, 1996) that would have variable spatial cognition and wayfinding and movement pattern impacts. Configuration characterizations allow for the distinction of slight and broad variations. In that sense, configuration is a more powerful spatial descriptor than composition. Linear configurational spatial characterizations, such as space syntax's axial map and segment map, have been used in inter-individual spatial cognition and wayfinding research as a complexity control dimension to help devise outdoor and 3D multi-level indoor wayfinding tasks (Hölscher, et al., 2006; 2011; 2012; Kuliga, et al., 2019; Lu & Ye, 2017; Dalton, 2003). Such descriptions are also well associated with aggregate pedestrian movement levels (Sharmin & Kamruzzaman, 2018). However, such representations have known limitations, as they treat multi-level vertical connections as an oversimplified two-dimensional encoding (Hillier & Hanson, 1984, p. 272; Hanson, 1998, pp. 53-54; Chang & Penn, 1998; Hölscher, et al., 2006; 2012). The improved space syntax segment representation and angular analysis (Hillier & Iida, 2005) remains un-operationalizable in multi-level 3D. There are calls for representations of 3D layout complexity to study inter-individual handling of verticality in wayfinding studies (Hölscher, et al., 2013; Kuliga, et al., 2019; Montello, 2007).

In this research, we propose to extend the principle of 2D road center line encoding (Turner, 2007), also proposed by Chang and Penn (1998), from 2D to 3D, using a 3D link/node medial center path for pedestrians representation. In a new software (Cooper & Chiaradia, 2020) we operationalize the representation to objectively measure 3D layout complexity by extending 2D angular analysis to 3D and to a novel 3D hybrid metric combining angular and metrical analysis (Montello, 2007). Central in Hong Kong is used as an extreme case study (Flyvbjerg, 2001). It is an urban setting including an entanglement of large scale vertical and horizontal, at ground, above and below ground, indoor and outdoor, public and private realms including a succession of large atriums and a range of seamlessly functioning vertical transitions (table S1, S2). The specificity of such transport oriented development developments is an emerging world-wide urban phenomenon called volumetric urbanism (Cui, et al., 2013; 2015; Cho, et al., 2015; Harris, 2015; Shelton, et al., 2011; McNeill, 2019;

Yoos & James, 2016; Mangin, et al., 2016). The complexity of these urban structures to a large part determines the complexity of wayfinding (Richter K, 2009) for millions of people daily. However, research still lacks appropriate 3D representation and analysis of such complex configurations (Hölscher, et al., 2013; Kuliga, et al., 2019). The aim of this study is to propose, test and compare novel representation, software and analysis to measure pedestrian layout complexity of emerging volumetric BE spatial configurations.

## Review of Relevant Studies

There are three prevailing motivations for using space syntax to analyze the impact of BE spatial structure complexity on navigation and wayfinding (Penn, 2003; Hillier, 1999; Zimring & Dalton, 2003; Hölscher, et al., 2012; Kuliga, et al., 2019; Carlson, et al., 2010):

1. Spatial linear representation: the axial line and its spatial disaggregation/derivation into segment map representing line of sight/movement, from which quantitative layout complexity measures such as topological 2D and 2.5D analysis are derived (Bafna, 2003; Haq & Zimring, 2003; Kuliga, et al., 2019). The more recent segment map enable angular and metrical analysis (Hillier & Iida, 2005).
2. From the linear representation, 2D or 2.5D quantitative configurational analyses of the spatial linear representation have a high association with the aggregate movement rates of pedestrians and vehicles, wayfinding decisions, and wayfinders' understanding of the spatial environment (Kuliga, et al., 2019), with unexplained variances often being associated with individual differences (Penn, 2003).
3. An intelligibility indicator, as a "local area effect", is defined as the correlation between the local measure of connectivity and global measures of integration. The high/low measure of the "inherent intelligibility" (Penn, 2003; Li & Klippel, 2016) of a spatial configuration is used as a predictor/explanation of the strong/weak association of space syntax analysis with movement rates in intelligible/"unintelligible", 2D, and 3D BE (Chang & Penn, 1998; Penn, 2003).

For a recent, extensive review and contextualization of the concepts, methods, and analyses of space syntax, see Rashid (2019) and Oliveira (2016).

### Spatial representation – axial line, segment line, and evolution to 3D

The axial map line representation, as the "the fewest and longest line" covering network with topological analysis is one of the original techniques of space syntax (Hillier & Hanson, 1984). It was developed to describe and characterize buildings and settlements and enable the explicit measurement of spatial configuration simplicity/complexity (Hillier, 1996; Hanson, 1998). Axial maps have been used to represent street networks, 2D outdoor BEs, 2D pedestrian paths, and 2.5D multi-level indoor buildings (Chang & Penn, 1998; Hölscher, et al., 2006; 2011; 2012). Early criticism of this method (Jiang & Claramunt, 2002; Ratti, 2004; Salheen & Forsyth, 2001) pointed out the relatively subjective nature of the encoding process and solution (Batty, 2004; Turner, et al., 2005), the lack of geometric, scale, and metrical considerations, the aggregated nature of the representation, its two-dimensionality (Pafka, et al., 2018), and the under determination of the axial line as the "line of sight" and/or line of movement (Marshall, 2005, p. 115; Emo, 2014; Steadman, 2004; Pafka, et al., 2018).

### Geometry, scale and metrical considerations, line of movement, line of sight

At the core of layout complexity measurement is the idea that the BE can be understood as a network of interconnected units. Complexity, initially conceptualized as a topological property of axial line and how axial line as units of analysis are interconnected to each other, has evolved with segment map, angular analysis, and metrical radius creating a geometric relationship with Euclidean metric conditioning in 2D and 2.5D. This allowed new "angular" analysis with a radius expressed either as the number of turns or steps (segments or metrical units). The combination of the segment map as a street representation using angular analytics and Euclidean radius has shown a higher level of association with pedestrian and vehicular flow (Hillier & Iida, 2005; Hillier, 2012). Axial and segment representations have been equally used (Sharmin & Kamruzzaman, 2018). A first hypothesis is that, given an operationalizable 3D representation, the Euclidean conditioned angular analysis can be extended to 3D.

The connectivity and continuity of the line of movement are entangled with sight, spatial cognition, and wayfinding heuristics but this does not necessarily equate "axial line" with line of sight (Turner, 2007; Marshall, 2005). It is hypothesized as suggested by Turner (2007) that the center line, the line of movement, is a valid representation/generalization of line of sight for pedestrian.

### Spatial aggregated nature of the axial/segment line representation of pedestrian path

Desyllas and Duxbury (2001), noting different pedestrian flow patterns on two sides of the street, proposed encoding the details of the pedestrian network as opposed to using aggregated encodings, such as axial lines or segment maps, that generalize the pedestrian paths of the two sides of street into one single line representation. Visual Graph Analysis (VGA) (Desyllas, et al., 2003; Turner, et al., 2001) was shown to be effective at capturing two sides of the street, pedestrian crossing configurations, and showing high association with pedestrian flow. However, empirical VGA and detailed pedestrian network studies are almost non-existent in the literature. A hypothesis is to extend the principle of the road center line 2D representation from vehicular movement to detailed pedestrian path encoding. This is effectively a road center line network linear referencing system used in transport geography (Figure 1a), and a standard data model representing a street network in transport (Garrison & Marble, 1962; Haggett & Chorley, 1969; ISO, 2011; FHWA, 2014; Marshall, 2016). These representations differ from that used by Turner (2007), as they do not decompose the curve into small straight segments. Instead, the curved link between two junctions (Figure 1a) is one spatial unit. Angular analyses is operationalized as angular change along a path is calculated cumulatively along links and through junctions (Cooper & Chiaradia, 2020).

The encoding principle follows a medial center path of the pedestrian network on both side of the street and the crossings, as shown in Figure 1b. This principle also follows pedestrian path encoding, not implemented in space syntax, proposed by Chang and Penn (1998, p. 525). Extending this encoding principle from indoor to outdoor should overcomes the inconsistencies arising with VGA (Ericson, et al., 2020; Zhang, et al., 2013) and the criticism of axial or segment line aggregated representation (Pafka, et al., 2018).

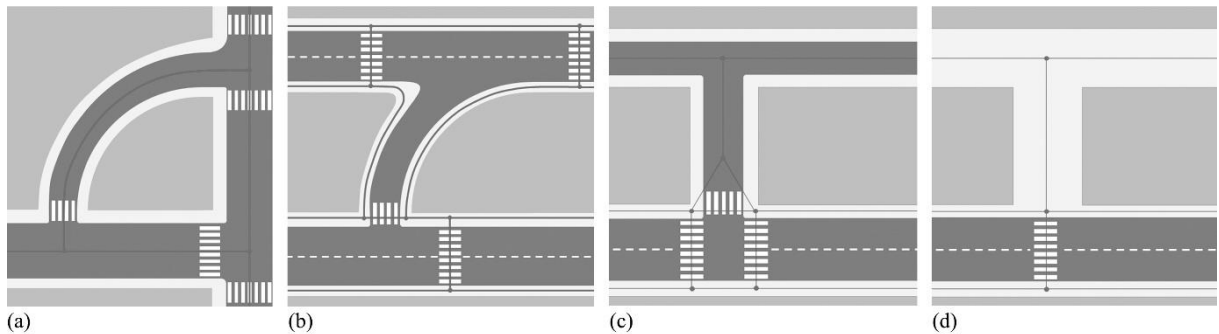

**Figure 1.** Outdoor pedestrian network encoding, link and node: **a**. 2D road center line encoding; **b.** desegregated 2D medial center path pedestrian network; **c. & d.** 2D hybrid medial pedestrian representation - **c.** link and node: residential streets with low traffic and no formalized crossing linked to a street with formalized crossings; **d**. link and node: pedestrianized streets linked to a street with formalized crossings.

### From 2D - 2.5D the axial/segment line to 3D

Chang and Penn (1998) and others (Kuliga, et al., 2019; Chang, 2002; Lu & Ye, 2017; Hölscher, et al., 2006; 2012) have highlighted the important role of vertical transition. There are calls for better encoding of vertical transition configurations to better understand their roles in multi-level navigation and wayfinding (Hölscher, et al., 2013; Lu & Ye, 2017; Montello, 2007). Theoretical extensions of space syntax to 3D have been proposed (Rashid, 2019, pp. 221-223), none of them have been empirically tested and operationalized in movement patterns and wayfinding studies. Indoor and outdoor multi-level pedestrian movement and wayfinding studies (Kuliga, et al., 2019; Chang & Penn, 1998; Hölscher, et al., 2012) using space syntax resort to an oversimplified linear representation of vertical transition configurations by using, on the same plane, an additional axial line for each connection between floors. This makes it at most a 2.5D representation. The 2.5D limitation makes it extremely difficult to operationalize the representation of a multi-level urban area that combines indoor and outdoor publicly accessible paths, such as in Chang and Penn (1998). Key publicly accessible indoor floors were not included in the pedestrian map. We hypothesize that the omission of interdependent, indoor-outdoor, publicly accessible pedestrian paths would impact negatively on correlation with movement patterns.

In spatial cognition and wayfinding studies, there is a debate as to whether or not cognitive maps are mainly 2D, or could be 3D, and whether or not users always expect identical floor layouts in multi-level buildings (Carlson, et al., 2010; Hölscher, et al., 2006; 2012). In Hong Kong, Lu and Ye's wayfinding study (2017) suggested that multi-level buildings can be integrated into a volumetric cognitive map and that the previous findings might be limited by the specific wayfinding task protocol used in a multi-level study (Berthoz & Thibault, 2013; Thibault, et al., 2013). While research as to whether vertical space is encoded equally or less accurately than horizontal space is still ongoing (Hinterecker, et al., 2018; Grieves, et al., 2020), to contribute to this debate meaningfully, the 2D hybrid representation (Figs. 1c & d) proposed by Cooper et al. (2019) is proposed to be deployed in 3D in order to encode topography, lift, stairway, ramp, or escalator, combining horizontal and vertical curvilinearity as shown in Figure 2 (Cooper & Chiaradia, 2020). These 3D encoding principles are a major change from the space syntax 2.5D linear network-based configurational approach to a complex 3D multi-level indoor and outdoor spatial structure. This 3D representation provides objective measures of body rotation between floors and level of correspondence floor by floor and across floors.

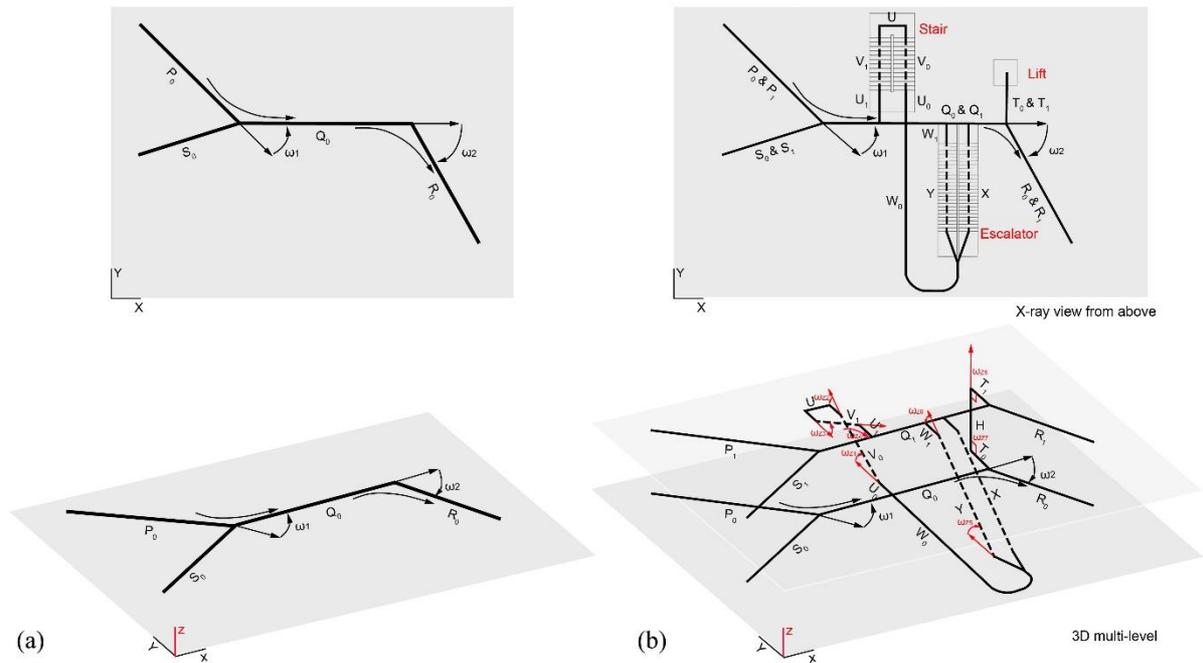

**Figure 2.** Pedestrian network encoding in 3D. **a**. road center line encoding and angular analysis principle (Turner, 2007) and **b**. 3D medial center path encoding and 3D angular analysis principle.

**Associations with pedestrian movement, vehicular and cycling movement patterns**

The strong association with aggregate observed pedestrian and vehicular movement distribution is a key motivation for wayfinding studies to use space syntax (Penn, 2003; Kuliga, et al., 2019; Farr, et al., 2012). In statistical terms, 60 to 80% of the variance in movement rates can be accounted for by measures of spatial configuration alone.

Although these associations between layout configuration and aggregate movement patterns provide no immediate understanding of individual cognitive processes, it is theorized that spatial cognition and navigational expertise, whether as walking, cycling or driving, elicit the deployment of cognitive processes and wayfinding heuristics that are also dependent on the layout complexity (Woollett & Maguire, 2011; Penn, 2003; Hillier & Iida, 2005; Hölscher, et al., 2012). Reciprocally, we hypothesize that the same configurational analysis using a different 2D representation, presenting similar associations with pedestrian, cyclist, and vehicular movement rates, is a plausible candidate to objectively describe BE complexity for wayfinding studies. The proposed representation displays similar or better association with pedestrian, cyclist and vehicular movement rates (Kang, 2017; Cooper & Chiaradia, 2015; Cooper, 2017; 2018; Wedderburn & Chiaradia, 2014; Jayasinghe, et al., 2019). Cooper et al. (2019), in the first longitudinal study, showed that the use of proposed representation enables predictive pedestrian movement after a major configuration change in a city center.

**Intelligibility – unintelligibility?**

The notion of intelligibility and its counterpart, unintelligibility, and their relation to spatial cognition is a widespread area of interest in space syntax literature (Penn, 2003; Dalton, et al., 2015; Haq, 2003). Legibility (Weisman, 1981) and intelligibility have been shown to be related (Long & Baran, 2012). A metric of legibility includes the connectivity metrics of layout, as interconnection density (O'Neill, 1991) which is fraught with the Modifiable Area Unit Problem (MAUP). Hillier (1996) exposed the concept of intelligibility as related to the ability to make inferences at a strategic location about the global spatial structure that lies beyond immediate visual local cues – "in which a picture of the whole urban system can be built up from its parts, and more specifically, from moving around from one part to another". These inferences are operationalized as an intelligibility index, the correlation of axial line local measure (Connectivity) to global measure (Integration Rn). High/low intelligibility is a good predictor of the correlation level (high/low) between space syntax analysis and pedestrian movement patterns (Penn, 2003). For example, 3D multi-level configurations in London, analyzed by Chang and Penn (1998) were deemed unintelligible thus explaining the result of the poor correlation between pedestrian movement patterns and space syntax metrics. However, despite some confounding issues with the operationalization of an intelligibility index (Zhang, et al., 2013; Long & Baran, 2012), it appears that the spatial representation and analysis used play an unaccounted major role in the level of correlation with the pedestrian movement pattern. Hillier and Iida (2005) show that the changes from axial to the derived segment line representations and topological to angular analyses vary the correlation with the pedestrian movement rate from -13 to +54% without a change of configuration, and thus no change in the intelligibility index. These variations cast doubt on the robustness and reliability of the intelligibility index.

The large scale studies of Coutrot et al. (2020; 2018) indicate that cognitive abilities for spatial navigation are clustered globally according to economic wealth and gender. Also, growing up in a city with a "complex" layout positively impacts navigational skills. These studies suggest that intelligibility will also vary not only with inter-individual but also with cross-cultural differences. Thus, an unintelligible layout in London might be intelligible in Hong Kong. This condition problematizes the construct of "inherent intelligibility" (Penn, 2003).

In summary, this study hypothesizes that as the definition and completeness of 3D representation increase we would found better association with movement patterns in large scale indoor and outdoor multi-level BE than 2.5D axial and segment maps, using the association with pedestrian movement patterns as the control. Building on this hypothesis, this study also examines how publicly accessible outdoor and indoor spaces are interdependent and affect associations with pedestrian movement patterns in such volumetric environments. We use an extreme case research strategy (Flyvbjerg, 2001). The case study area selected has an overall intelelligibility close to zero.

## Datasets and Methods

### The Study Area

With over 7.4 million people and a built-up land area covering approximately 25% of its territory, Hong Kong has one of the highest population densities in the world. Walking is the main way of moving between public transport modes that are used by 90% of the population. Hong Kong has also adapted walking to a typhoon-prone, sub-tropical climate, and the city's hilly terrain. It has evolved into a complex volumetric mode of development (Shelton, et al., 2011), broadly defined as integrating multiple modes of transport with vertically stacked mixed use, through walking below, at, and above ground, combining outdoors and indoors, and mingling public, quasi-public, and private ownerships that are publicly accessible and seamless. Such urban development can be found in other cities around mass transit urban rail stations (Yoos & James, 2016; Mangin, et al., 2016; Samant, 2019; Chang & Penn, 1998).

The study area is defined as an 800 m catchment area (2.2 x 1.6 km) from the entrances/exits of two linked MTR stations: Central and Hong Kong stations. It excludes the very constrained inner-urban rail station's paying area. Workplace density within the study area, one of the highest in the world, ranges from 1,140 to 2,990 jobs per hectare, while the residential density is relatively low for Hong Kong, ranging from 12 to 465/ha. These differentiated ranges are a typical functional signature of a Central Business District and implies that most users are familiar with the BE configuration. The study area also attracts many tourists who might mostly be first time visitors.

The multi-level outdoor covered walkway network of the Central area is 7.8 km long. It has been developed piecemeal by the Hong Kong Government and various private developers since the 1970s (Tan & Xue, 2014). This multi-level pedestrian network extends horizontally and vertically, playing an essential role in connecting dense, multi-level, publicly accessible mixed-use areas (Figure 3). This complex urban environment has attracted various descriptive studies (Frampton, et al., 2012; Xue, et al., 2012; Xue C, 2016; Zhang, 2017; Sun, et al., 2019).

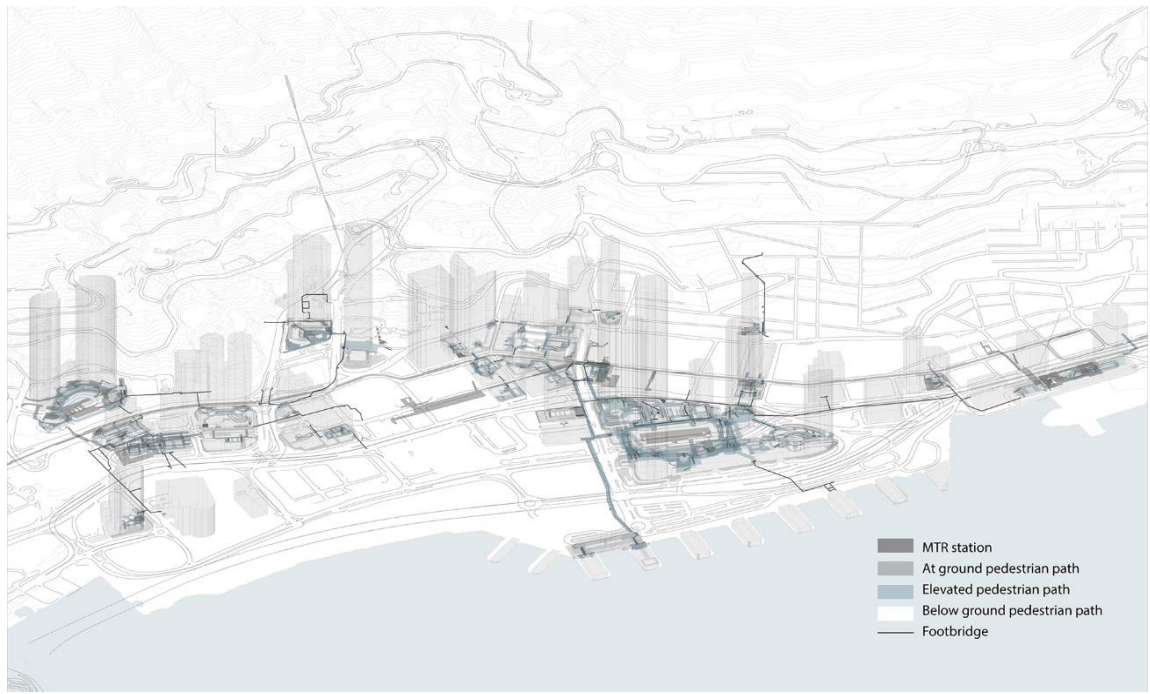

**Figure 3.** Hong Kong, between Sheung Wan (right) and Admiralty (left), Central (middle) volumetric mega-structure.

### Pedestrian Data Collection

Pedestrian movement pattern data was gathered in the Central area as a dependent variable and used as a quantitative control test for the objective BE representations and analysis as independent variables. Details of the collection methods are in the SM01. Table S2 & Figure S1 shows the descriptive statistics for all observed cordon counts. The pedestrian flows on the skywalk are higher than those at ground level. Figure S2 presents the correlations between all-day average flow rates and each period. The average of 6 time periods throughout the day is a good representation of each period.

### Pedestrian Network Representations

In this study, four spatial representations are compared with different level of 3D definition and level of completeness, yielding seven representations. Detailed descriptions are given in the SM02. A graphic summary is shown in Figure 4.

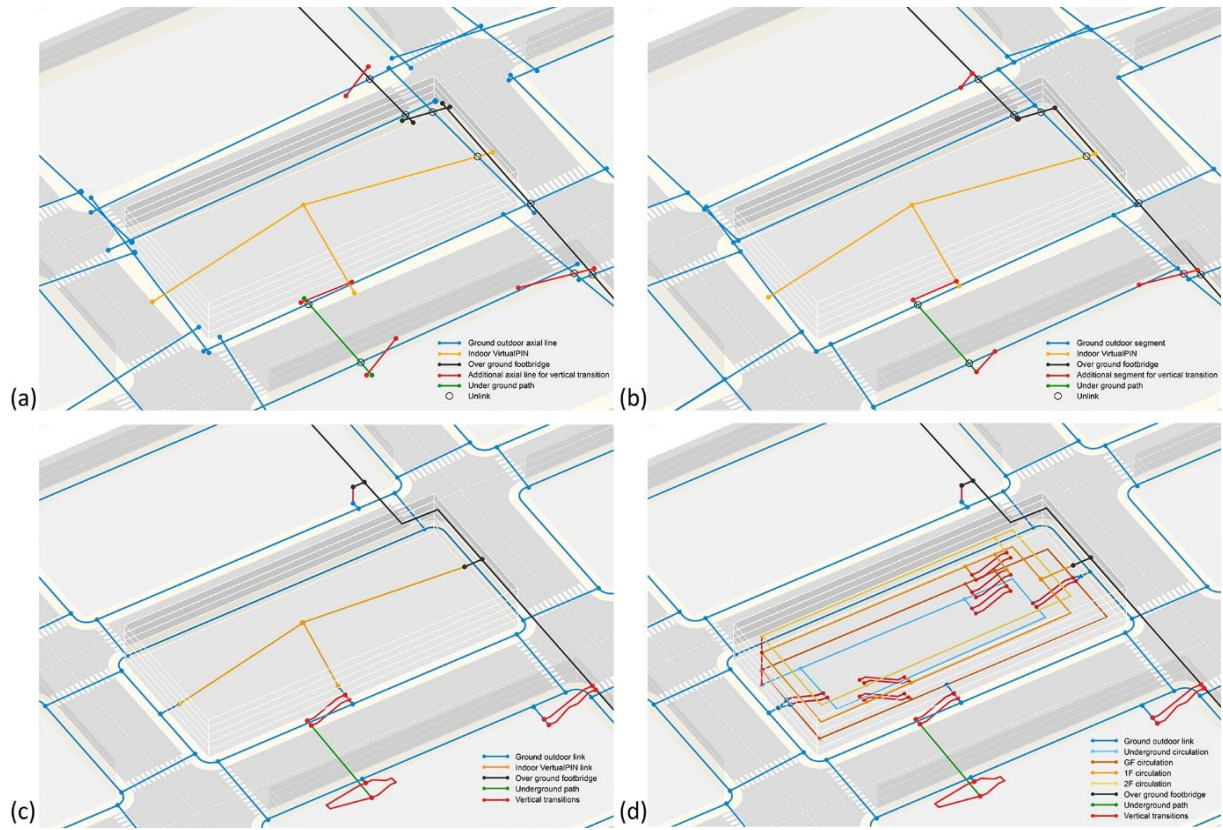

**Figure 4.** Four spatial representations of pedestrian network:

*space syntax, 2.5D-depthmap software:* **a.** 2.5D Axial map - outdoor at grade, skywalk, below ground indoor, unlinks, PIN in orange, and added line as vertical transition; **b.** 2.5D Segment map - derived from Axial map, it follows the same principles, PIN in orange;

*3D-sDNA software:* **c.** 3D Path-center line map outdoor at grade, below ground, skywalk, escalator, lift, ramp, stairway, and PIN in orange; **d.** 3D Path-center line map outdoor and full multi-level indoor with vertical transition: escalator, lift, ramp, and stairway.

Table S3 compares the descriptive statistics of the axial, segment, 3D outdoor path-center line, and complete 3D indoor-outdoor path-center line maps. The increased number of links and decreased average link length from the axial map to the segment map illustrate the aggregated nature of the axial map (Ratti, 2004).

### Analytical Metrics: Centrality Independent Variables and Software

Two measures of pedestrian network accessibility and flow potential were used in this study: closeness centrality and betweenness centrality (Hillier & Iida, 2005; Cooper & Chiaradia, 2020). In space syntax software, closeness centrality is called 'integration' or 'mean depth', and betweenness centrality is called 'choice' (Hillier & Hanson, 1984).

**Axial and segment maps** are processed with open source DepthmapX (Varoudis, 2012).

**Path-center line 3D maps** are analyzed with the open-source 3D sDNA toolbox for ArcGIS and QGIS (Chiaradia, et al., 2014). Figure 5b illustrates how a 3D spherical Euclidean radius operates in a 3D network. Closeness and betweenness centrality are used (Cooper & Chiaradia, 2020).

*Radius,* according to the Travel Characteristics Survey 2011 (Transport Department, 2014), over 75% of Hong Kong residents walked up to 5 minutes, about 400 m or less, from their trip origin to access either a mechanized transport mode or their trip destination. For axial maps, choice and integration are calculated with radii from R3 to R10. Similarly, for segment maps, choice and integration are calculated for every 100 m radii within the walking distance range of 200 to 500 m.

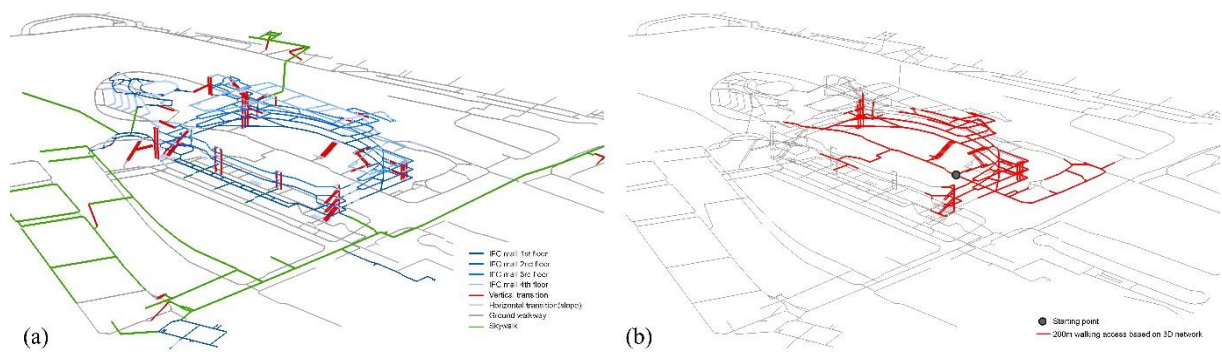

**Figure 5. a.** 3D indoor-outdoor network of IFC mall; **b.** Illustration of 3D spherical 400 m radius.

### Hybrid Distance Metric

sDNA as DepthmapX can deploy different analytical metrics such as angular (most direct path), Euclidean (shortest path), and topological (least turn path). sDNA provides a novel 'hybrid' metric (shortest and most direct). In a recent study, Shatu et al*.* (2019) reported that pedestrians tend to minimize the horizontal directional changes and path lengths if they can and that 28% of the chosen routes satisfied both distance and direction criteria. Berthoz et al. (1999) indicated that distance and direction might be encoded separately yet interacting. Spatial configuration can be characterized by the number of paths that are more or less the shortest but also the most geometrically direct (Zhang, et al., 2015). sDNA allows the mixing of both the shortest and least angular changes in one metric to explore their relative interaction. The hybrid distance metric is defined along the link in Equation (3) and for the junction in Equation (4).

$$\text{distanceforlink} = a \times ang + (1-a) \times euc \qquad (1)$$
$$\text{distancefornode} = a \times ang \qquad (2)$$

The weighting constant *a* specifies the relative importance of angular distance over Euclidean distance, where *a* = 1 represents pure angular, and *a* = 0 represents pure Euclidean. *ang* is the cumulative angular change along a link or angular change at a junction. *euc* is the Euclidean distance measured along the path. We tested the hybrid metric by varying the angular-Euclidean ratio to investigate different weighting: 0:1 (pure angular), 1:2, 1:1, 2:1, 1:0 (pure Euclidean).

### Results

To test the relationship between spatial configuration representations at different radii (3 to 7 for axial maps, 200 to 500 for segment and path-center line maps) and observed average daily pedestrian movement, a bivariate correlation was adopted for all the seven spatial representations (Figure S3). The R-square correlation values between pedestrian movement pattern distribution and centrality measures are shown in Table 1.

The results suggest that the pedestrian movement pattern's association with the measure of layout complexity in volumetric BE is significantly affected by the level of definition and completeness of the pedestrian path layout in a 3D representation (see SM03). Association with a movement pattern increases steadily as the representation increases toward high 3D definition and completeness.

Table 1. R-square correlations between Ln pedestrian movement and Ln betweenness/Ln Closeness measures for 4 different representations **(N=33)**.

| | | Angular/ Euclidean metric | Axial map 2.5D | Segment map 2.5D | 3D Outdoor | Axial map 2.5D | Segment map 2.5D | 3D Outdoor | 3D Outdoor & Indoor |
|---|---|---|---|---|---|---|---|---|---|
| | | | | Without PIN | | | With PIN | | |
| Betweenness | Topological | NA | **0.288(R6)**** | NA | 0.046(R400) | **0.345(R7)**** | NA | 0.177 (R500)* | 0.177 (R400)* |
| | Angular | 1:0 | NA | **0.217 (R600)**** | 0.052 (R500) | NA | **0.308 (R600)**** | 0.252 (R600)** | **0.489 (R500)**** |
| | Euclidean | 0:1 | NA | 0.057 (R600) | 0.096 (R500) | NA | 0.204 (R500)** | **0.284 (R500)**** | **0.507 (R500)**** |
| | Hybrid | 1:1 | NA | NA | 0.011 (R400) | NA | NA | 0.258 (R600)** | 0.516 (R500)** |
| | Hybrid | 1:2 | NA | NA | - | NA | NA | - | **0.524 (R500)**** |
| | Hybrid | 2:1 | NA | NA | - | NA | NA | - | 0.501 (R500)** |
| Closeness | Topological | NA | **0.173** (R9)* | NA | 0.403(R600)** | **0.384** (R8)** | NA | 0.145 (R600)* | 0.085 (R400) |
| | Angular | 1:0 | NA | **0.261** (R600)** | **0.537** (R400)** | NA | **0.556** (R500)** | **0.570** (R600)** | **0.718 (R500)**** |
| | Euclidean | 0:1 | NA | 0.235 (R500)** | 0.047 (R500)** | NA | 0.077 (R500) | 0.020 (R600) | **0.180 (R600)*** |
| | Hybrid | 1:1 | NA | NA | 0.423 (R400)** | NA | NA | **0.591** (R500)** | 0.677 (R500)** |
| | Hybrid | 1:2 | NA | NA | - | NA | NA | - | 0.654 (R500)** |
| | Hybrid | 2:1 | NA | NA | - | NA | NA | - | **0.679 (R500)**** |

*Note.* Axial & segment maps were computed by Space Syntax software (DepthmapX), outdoor only & indoor-outdoor path-center line maps were computed by Spatial Design Network Analysis (sDNA). Betweenness and closeness measures were analyzed at radius R5, 6, and 7 for axial map, and radius 400, 500, and 600 m for segment and path-center line maps, numbers in round parenthesis indicate the best radius.

*\* p≤0.05 level. \*\* p≤0.01 level.*

Table 2 shows cross-correlations among four metrics of betweenness measures for a 3D indoor & outdoor pedestrian network. The strongest association is between hybrid (1:1) and Euclidean betweenness, with $R^2 = 0.989$. The association between hybrid and angular betweenness is $R^2 = 0.877$. These results demonstrate that a hybrid metric, combining angular and Euclidean metrics, enables a better understanding of their interactions.

Table 2. Matrix of R-square correlation (Radius at 500 m) between betweenness measures of 3D indoor-outdoor pedestrian network **(N=33)**.

|            | Angular    | Euclidean  | Topologic  | Hybrid |
|------------|------------|------------|------------|--------|
| Angular    | 1          |            |            |        |
| Euclidean  | 0.834**    | 1          |            |        |
| Topologic  | 0.570**    | 0.694**    | 1          |        |
| Hybrid     | 0.877**    | 0.989**    | 0.597**    | 1      |

*\*\* p≤0.01 level.*

We varied the pedestrian survey methods using one weekday video survey and 3 days video survey to further ascertain the robustness of the findings. The results hold, see details in SM04.

## Discussion and Conclusion

In this study, we show for the first time how angular analysis, as an objective measure of pedestrian layout complexity and index of wayfinding difficulty (Carlson, et al., 2010), can be innovatively deployed in 3D by changing representation. Seven representations were comparably analyzed by using angular path preference analysis across junctions and links in a large scale "unintelligible" outdoor and indoor BE in Central, Hong Kong. The key innovation is that a link, as a spatial unit of analysis, combines horizontal and vertical curvilinearity, making it truly 3D (Figure 2). We demonstrated that this 3D representation can also be used to represent the details of the outdoor pedestrian pathways on both sides of streets, including crossings, in a topographically rich study area with extensive multi-level outdoor and indoor pedestrian paths, i.e., in volumetric urbanism (Shelton, et al., 2011; McNeill, 2019).

The axial, segment, and 3D maps were also comparably analyzed. We found that the association with movement patterns for the axial map was moderate in contrast to low found in a previous study (Chang & Penn, 1998). This may be because Hong Kong has a complex spatial culture (Coutrot, et al., 2018; 2020). Association with the pedestrian movement pattern increased to a high level ($R^2 \approx 0.72$) as the spatial level of realism increased from 2.5D axial to 2.5D segment, and from 3D outdoor, with and without a PIN, to a fully detailed 3D outdoor and indoor representation. In empirical studies of Barbican and South Bank in London, Chang and Penn (1998) attributed negligible correlations between the logarithm of pedestrian movement rates and integration values to the weak correlation between local and global measures, or the lack of "intelligibility" of the BE (0.21 in Barbican and 0.41 in South Bank). In this study area, intelligibility is almost null; the R-square correlation between connectivity and global integration is only 0.09 for the outdoor network with PIN. However, it appears that the predictability of a system is not related to the "intelligibility index" of an area. The success of this improved 3D full outdoor and indoor representation, combined with angular and hybrid analytics, foster an improved network-based configurational approach to the analysis of morphological complexity found in volumetric urbanism. These improved representations and analysis still stress the importance of configuration, geometry, Euclidean distance and enable the measure of complexity, degree of correspondence, direct measure of rotation, and overlap between floor to floor plans (Carlson, et al., 2010).

The implementation of a hybrid analysis, conditioned by the Euclidean radius, combined the angular and Euclidean versions of closeness and betweenness (integration & choice in space syntax). Given that angular and Euclidean distance appear to influence wayfinding, their combination and relative weighting reflect the cognitive shortest paths for different levels of knowledge of the system, conditioned on the level of overlap that may exist between the most direct and shortest paths (Zhang, et al., 2015; Shatu, et al., 2019).

Volumetric urbanism, transport-oriented development, and large-scale multi-level BE create new cognitive, wayfinding and urban planning and design challenges due to their spatial scale and complexity. These complexities can increase due to incremental additive development over time, such as in Central, Hong Kong, as opposed to "all at once" design, as for example in Sha Tin, Hong Kong or the Barbican in London. Both design approach have their issues, such as spatial complexity and wayfinding difficulty that were lacking robust analytics. Consequently, this study's results suggest that layout complexity analysis of pedestrian network map should include complete 3D outdoor and indoor pedestrian paths.

Given the high level of complexity and correlation of the study area and movement patterns, it appears that the understanding of local spatial culture should be given prominence. Therefore, the direct import of best practice between different spatial cultures comes into question. In that sense, the comparison of pedestrian patterns in New York, Manhattan, and Hong Kong, because they have similar densities is not warranted given their spatial culture differences, flat regular grid vs topography rich, deformed grid and multi-level. It raises an interesting challenge to designers, researcher and policymakers accustomed to liberal borrowing from various spatial cultural settings and thus the need to better understand the evolving plasticity or not of spatial culture(s).

Standard 3D line representations support the inclusion of physical and more intangible attributes along configurational information of the BE. They also enable linear referencing of semantic attribute along movement lines or other BE attributes in 3D.

There are several limitations to this study. One of the limitations is that it is a single extreme layout complexity case study design, meaning that traditionally its findings may not have high external validity. Yet, as Flyvbjerg (2001) pointed out that extreme case, such as is Central in Hong Kong in term of spatial configuration complexity, reflects the highest status with respect to the variable being measured, the layout complexity/unintelligibility, and thus can be presented to capture the extreme extent of the diversity that can be observed in the reference class distribution, the range of complexities of 2D and 3D BE of other parts of Hong Kong and other cities.

Also, only weekday pedestrian volumes were examined, so an extension of this study would be to compare weekday and weekend pedestrian movement patterns. There is also no pedestrian flow measurement indoor due to private ownership that would have been beneficial to include if available. Further investigation is also needed on the overlapping relationship between the shortest angular path and the shortest Euclidean path in a complex 3D BE, using hybrid metrics. Moreover, further work could be conducted within Carlson et al.'s integrative framework (2010). The concept of the "inherent intelligibility" of a BE needs to be reconsidered when the representation itself impacts so much on the results. Neuroscience perspectives provide salient BE features and configurations that could be tested to better inform spatial design legibility-intelligibility (Jeffery, 2019). Given the plasticity of the cognitive processes and the varieties of individual wayfinding capacities and strategies deployed, future research on the dynamics of individual and cultural differences that interplay with life-long change should be conducted to improve the understanding of the relationship between the BE, user characteristics, and learning period.

The associations between pedestrian movement patterns and some spatial variables, including pathway width and distance to the closest MTR station, were found to be insignificant in this case. However, other known variables (Cooper, et al., 2019) such as density, diversity and destination inter-accessibility were omitted to focus on comparing the impact of varying pedestrian network representations on configuration variables against the same set of pedestrian movement pattern records.